\documentclass[12pt,titlepage]{utarticle}

\DeclareMathSymbol{\square}{\mathord}{AMSa}{"03}
\DeclareMathSymbol{\rightsquigarrow}{\mathrel}{AMSa}{"20}
\renewcommand{\Box}{\square}
\renewcommand{\leadsto}{\rightsquigarrow}

\newdimen\tableauside\tableauside=1.0ex
\newdimen\tableaurule\tableaurule=0.4pt
\newdimen\tableaustep
\def\phantomhrule#1{\hbox{\vbox to0pt{\hrule height\tableaurule width#1\vss}}}
\def\phantomvrule#1{\vbox{\hbox to0pt{\vrule width\tableaurule height#1\hss}}}
\def\sqr{\vbox{%
  \phantomhrule\tableaustep
  \hbox{\phantomvrule\tableaustep\kern\tableaustep\phantomvrule\tableaustep}%
  \hbox{\vbox{\phantomhrule\tableauside}\kern-\tableaurule}}}
\def\squares#1{\hbox{\count0=#1\noindent\loop\sqr
  \advance\count0 by-1 \ifnum\count0>0\repeat}}
\def\tableau#1{\vcenter{\offinterlineskip
  \tableaustep=\tableauside\advance\tableaustep by-\tableaurule
  \kern\normallineskip\hbox
    {\kern\normallineskip\vbox
      {\gettableau#1 0 }%
     \kern\normallineskip\kern\tableaurule}%
  \kern\normallineskip\kern\tableaurule}}
\def\gettableau#1 {\ifnum#1=0\let\next=\null\else
  \squares{#1}\let\next=\gettableau\fi\next}

\begin{document}

\preprint{
 UTTG--20--96\\
 {\tt hep-th/9611088}\\
}

\title{$N=1$ Dualities for Exceptional Gauge Groups\\
		and Quantum Global Symmetries}
\author{Jacques Distler and Andreas Karch
 \thanks{Work supported in part by NSF Grant PHY9511632, the Robert A.~Welch
    Foundation, an Alfred P.~Sloan Foundation Fellowship and the DAAD.
 }
 \oneaddress{
  Theory Group\\
  Department of Physics\\
  University of Texas\\
  Austin, TX 78712, USA\\
  {~}\\
  \email{distler@golem.ph.utexas.edu}
  \email{akarch@physics.utexas.edu}
 }
}
\date{November 11, 1996}

\Abstract{
We discuss our attempts to generalize the known examples of dualities in $N=1$
supersymmetric gauge theories to
exceptional gauge groups. We derive some dual pairs from known
examples connected to exceptional groups and find an interesting phenomenon:
sometimes
the full global symmetry
is ``hidden'' on the magnetic side. It is not realized as a symmetry on the
fundamental fields in the Lagrangian. Rather, it emerges as a symmetry of the
quantum theory. We then focus on an approach based on self-dual models. We
construct duals for some very special matter content of $E_ 7$,
$E_6$ and $F_4$. Again we find that the full global
symmetry is not realized on the fundamental fields.
}

\maketitle

\section{Introduction}

The past few years have seen remarkable progress in the study of
supersymmetric gauge theories.
Most of this progress is due to the phenomenon of duality: two seemingly
different theories are shown to describe the same infrared physics. This
is a symmetry of the full theories, after all the non-perturbative effects
have been taken into account.
 Weak and strong coupling get interchanged.
This allows one to obtain results in strongly coupled theories that
were inaccessible before, once the duality is established.

Using these new symmetries lead to beautiful results in the last few years,
especially $N=2$ \cite{witten} and $N=4$ \cite{olive} supersymmetric gauge
theories. N.~Seiberg has shown in \cite{seiberg1,seiberg2,seiberg3}
that for $N=1$ theories a somewhat weaker version
of duality still holds: for certain values of the adjustable parameters two
theories different in the ultraviolet flow to the same infrared fixed
point. That means for an observer only testing the low energy regime, there
is no way to distinguish the two theories.

Seiberg did his original studies on $SU$ and $SO$ gauge theories with
matter only in the fundamental representation. Since then many new examples
have been found with more complicated matter content and different groups
\cite{intrilligator,pou7,
pou8, pou10, pouasy, berkooz, luty,sp}.
Despite all the effort that has gone in
finding new dualities, we are still lacking a dual description for some
of the most interesting theories: those with exceptional gauge groups. These
theories appear over and over again in connection with issues like string
phenomenology, dynamical supersymmetry breaking and SUSY grand unification.

In this paper, we describe our efforts to find dual descriptions for $N=1$
gauge theories with exceptional gauge groups and an interesting phenomenon
we discovered during our analysis: in some dual pairs the full global
symmetry is not visible in terms of the fundamental fields but only appears
as a symmetry of the effective theory in the far infrared. To use a
term which has become familiar in the string context, it is a \emph{quantum}
symmetry, not present as a symmetry of the Lagrangian, but which appear only
as a symmetry of the quantum theory.

We used two different approaches to the problem. First we tried to
generalize the dualities of Pouliot and Strassler \cite{pou7,pou8,pou10}
between $Spin(7,8,10)$ and $SU$ groups with a symmetric tensor.  These $Spin$
groups are reached when one higgses exceptional groups, so one might hope
that duals for exceptional groups should look  somewhat similar.

In the end, our second approach proved to be more fruitful. In \cite{leigh}
the special importance of self-dual models was pointed out. Ramond
found \cite{ramond} that $E_6$ with 6 flavors is self-dual. Similarly
we argue that $E_7$ with 4 flavors is self-dual. Starting from
those two theories we generate some new dual pairs by going
along flat directions and perturbing the theories with mass terms.
Again we find the global symmetries realized as quantum symmetries of the
magnetic theory.

In Section 2 we will review the $Spin(10)$ duality of \cite{pou10} which
we used as a starting point for the first part of our analysis.
In Section 3 we will discuss our close analysis of these $Spin
\leftrightarrow SU$ dualities. We will encounter for the first time that
the full global
symmetry acts in a very funny way on the magnetic side: it combines
fundamental fields and composite fields into multiplets of the
full global symmetry.

In section 4 we will use a self-dual model obtained by Ramond \cite{ramond}
for $E_6$ and a new self-dual model for $E_7$ to derive some
new dual pairs with exceptional gauge groups. In particular
duals are found for:

\begin{itemize}
\item $E_7$ with 4 fundamentals
\item $E_6$ with 3 fundamentals and 3 antifundamentals
\item $F_4$ with 5 fundamentals
\item $F_4$ with 4 fundamentals
\end{itemize}

In Section 5 we will give our conclusions.

\section{Duality in $Spin(10)$ with a Spinor}

Last year Pouliot and Strassler found dualities between
 $Spin(7,8,10)$ \cite{pou7,pou8,pou10} and
$SU$ groups with a symmetric tensor representation. For the $Spin(7)$
and $Spin(8)$ case the situation is really strange, since these dualities
map the non-chiral electric theory to a chiral magnetic theory. Since
we used their $Spin(10)$ example as a starting point we'd like
to review this duality. The other two examples have a
very similar structure.

As usual, this duality is not proven, but the conjecture is based
on some solid evidence: the global symmetries are the same, it satisfies
the 't Hooft anomaly matching conditions, and the gauge invariant operators
match. Under perturbations the duals
flow to new consistent dual pairs, as we will show in the next section.
Under a very special perturbation (giving a vev to a 10 of $Spin(10)$ and
then writing down a mass term for the spinor), they are able to connect
their dual to the
basic $SO$ duals of Intriligator  and Seiberg \cite{seiberg3}.

The electric theory of this dual pair is $Spin(10)$ with one spinor and
an arbitrary number of vectors.
If $n$, the number of additional vectors, is greater or equal to 7, the theory
is in its non-abelian Coulomb phase.  Pouliot
and Strassler established the following duality for this case:

\begin{center}

\begin{tabular}{|c||c||c|c|c|}
\hline
&Gauge Group&\multicolumn{3}{c|}{Global Symmetries}\\
\cline{2-5}
&$Spin(10)$&$SU(n)$&$U(1)$&$U(1)_R$\\
\hline
\hline
$Q \rule{0cm}{.48cm}$&10&$n$&$-1$&$1-\frac{8}{n+2}$\\
\hline
$S \rule{0cm}{.48cm}$&16&1&$\frac{n}{2}$&$1-\frac{8}{n+2}$\\
\hline
\end{tabular}
\par\medskip
$\updownarrow$
\par\medskip
\begin{tabular}{|c||c||c|c|c|}
\hline
&Gauge Group&\multicolumn{3}{c|}{Global Symmetries}\\
\cline{2-5}
&$SU(n-5)$&$SU(n)$&$U(1)$&$U(1)_R$\\
\hline
\hline
$q
\rule{0cm}{.48cm}$&$\overline{n-5}$&$\bar{n}$&$1$&$\frac{8}{n+2}-\frac{1}{n-5}$\\
\hline
$q\prime \rule{0cm}{.48cm}$&$n-5$&1&$-n$&$-1+\frac{16}{n+2}+\frac{1}{n-5}$\\
\hline
$s \rule{0cm}{.48cm}$&$\Box\!\Box$&1&0&$\frac{2}{n-5}$\\
\hline
\hline
$Y \rule{0cm}{.48cm}$&1&$n$&$n-1$&$3-\frac{24}{n-5}$\\
\hline
$M \rule{0cm}{.48cm}$&1&$\Box\!\Box$ &$-2$&$2-\frac{16}{n-5}$\\
\hline
\end{tabular}
\par\nopagebreak\medskip
with: $W= Mqsq+\det s+Yqq\prime$.
\end{center}

The gauge invariant polynomials get mapped as follows (
all gauge invariants  contracted with a $\delta$ are referred to as mesons
whereas baryons are
contracted with an $\epsilon$):

Mesons:
\begin{eqnarray*}
 \begin{array}{ccc}
QQ &\leftrightarrow& M\\
(SS)_s Q^5_a &\leftrightarrow& q^{n-5} \\
 \end{array}
&\hspace{1cm}&
 \begin{array}{ccc}
(SS)_s Q &\leftrightarrow&  Y  \\
 \end{array}
\\
\end{eqnarray*}

Baryons:
\begin{eqnarray*}
 i=0,1,2\\
 Q_a^{10-2i} W^i &\leftrightarrow& q^{n+2i-10} s^{n-8+i} W^{2-i} q\prime\\
 Q_a^{9-2i}  W^i (SS)_s &\leftrightarrow& q^{n+2i-9} s^{n-7+i} W^{2-i} \\
\end{eqnarray*}

Even though this duality was established as a symmetry between two
theories in an interacting non-abelian Coulomb phase ($N \geq 7$) one can
extrapolate to the case $n=6$. The duality maps to theories of singlets.
This is similar to what one does in SUSY QCD for $N_C+1 < N_F \leq
\frac{3}{2} N_c$. The electric theory leaves the non-abelian Coulomb phase.
One can
easily see this from the unitarity bound provided by the superconformal
algebra. The R-charge of any operator becomes related to its scaling dimension
\begin{eqnarray*}
D \geq \frac{3}{2} |R|
\end{eqnarray*}
the bound being saturated for chiral operators. A violation of
this bound signals a breakdown of conformal symmetry. To describe the
physics in this regime, one extrapolates the conjectured duality.
In SUSY QCD the magnetic theory is free and one hence concludes that
the electric theory confines, the composite fields being subject to
the new magnetic gauge dynamics. In the present case the magnetic theory
is also a theory of singlets. To maintain the notion of exchange of weak
and strong
coupling one should look at this case as a free magnetic phase with trivial
gauge group.

As a nice application
Pouliot and Strassler use this duality to reconfirm the known result
\cite{oldone} that $Spin(10)$ with just one spinor dynamically breaks
supersymmetry by studying the theory in the presence of mass terms.

\section{More on $Spin \leftrightarrow SU$ Dualities}

\subsection{Derived Dualities}

The three theories studied by Pouliot and Strassler \cite{pou7,pou8,pou10},
$Spin(7)$ with $m$ spinors,
$Spin(8)$ with 1 spinor and $n$ vectors and
$Spin(10)$ with 1 spinor and $n$ vectors
are connected to a wide variety of other very interesting gauge theories
via the Higgs mechanism,
including those we set out to solve, the exceptional ones:

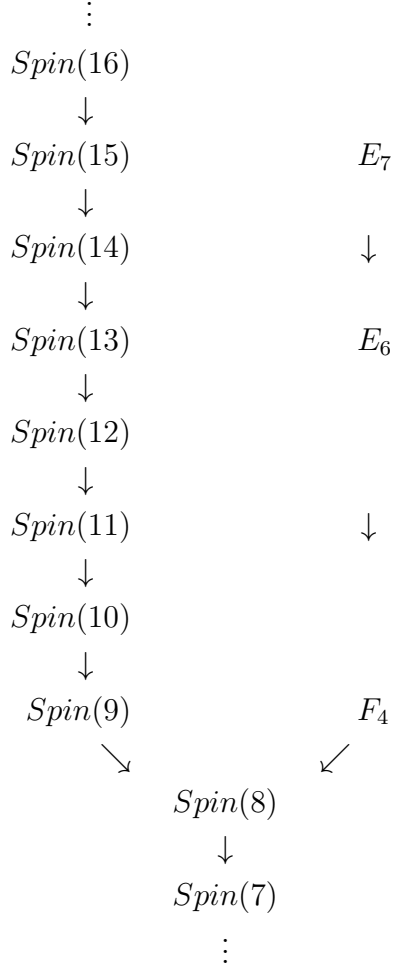
\begin{eqnarray*}
\vdots \hspace{0.5cm} &\hspace{1.8cm}&\\
Spin(16)&\hspace{1.8cm}&\\
\downarrow \hspace{0.5cm}&&\\
Spin(15)&\hspace{1.8cm}& \hspace{0.5cm} E_7\\
\downarrow \hspace{0.5cm}&&\\
Spin(14)&\hspace{1.8cm}& \hspace{0.55cm} \downarrow\\
\downarrow \hspace{0.5cm}&&\\
Spin(13)&\hspace{1.8cm}& \hspace{0.5cm} E_6\\
\downarrow \hspace{0.5cm}&&\\
Spin(12)&\hspace{1.8cm}&\\
\downarrow \hspace{0.5cm}&&\\
Spin(11)&\hspace{1.8cm}& \hspace{0.55cm} \downarrow\\
\downarrow \hspace{0.5cm}&&\\
Spin(10)&\hspace{1.8cm}&\\
\downarrow \hspace{0.5cm}&&\\
Spin(9)&\hspace{1cm}& \hspace{0.5cm} F_4\\
\searrow&&\swarrow\\
&Spin(8)&\\
&\downarrow&\\
&Spin(7)&\\
&\vdots&\\
\end{eqnarray*}

The arrows indicate higgsing of the gauge group by giving a vev to
a field transforming in the fundamental representations. For the
exceptional groups this procedure is not unambiguous: one can give
a fundamental a vev in different ways, resulting in different
unbroken subgroups. Or put another way, a fundamental field may
have more than one flat direction associated with it. Since
flat directions are parametrized by gauge invariant polynomials,
one can distinguish between the different possibilities by considering which
gauge invariant combination of fields gets a vev. The above diagram
reflects a vev to the invariant contracted with the symmetric
invariants special to the exceptional groups: $d^{\alpha \beta \gamma \delta}$
for $E_7$ $d^{\alpha \beta \gamma}$ for $E_6$ and $F_4$. For $F_4$ there
is another possibility by contracting with $\delta^{\alpha \beta}$, the
corresponding vev
breaking the group only to $Spin(9)$. For $E_6$
there is an additional invariant involving a fundamental and an
antifundamental,
$\delta^{\alpha}_{\beta}$, and  for $E_7$ there is an additional invariant
antisymmetric in two fundamentals, $f^{\alpha \beta}$. Their vevs leave
unbroken, respectively, a $Spin(10)$ or a $Spin(11)$ subgroup.

In our attempt to generate a dual picture for the above diagram we tried
some obvious generalizations of Pouliot's duality. Those models had some
very nice features, like matching of 't Hooft anomalies and of {\it some}
gauge invariant operators, but in the end they all proved to be inconsistent.
The ``troublemakers'' were in the first place the gauge invariants
symmetric in the flavor indices.

To get a better understanding, where the difficulties lie and how to
deal with those strange dualities, we studied the continuation of
the diagram to smaller $Spin$ groups:

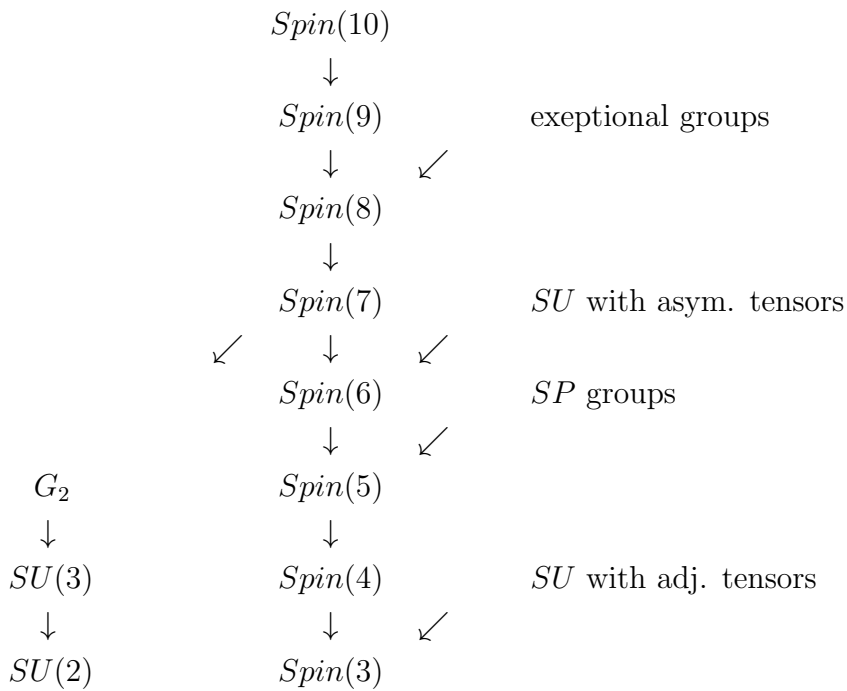
\begin{eqnarray*}
&Spin(10)&\\
&\downarrow&\\
&Spin(9)& \hspace{1.5cm} \mbox{exeptional groups}\\
&\downarrow&\swarrow\\
&Spin(8)&\\
&\downarrow&\\
&Spin(7)& \hspace{1.5cm} \mbox{$SU$ with asym. tensors}\\
\swarrow&\downarrow&\swarrow\\
&Spin(6)& \hspace{1.5cm} \mbox{$SP$ groups}\\
&\downarrow&\swarrow\\
G_2 \hspace{2.3cm}&Spin(5)&\\
\downarrow \hspace{2.5cm} &\downarrow&\\
SU(3) \hspace{2cm}&Spin(4)& \hspace{1.5cm} \mbox{$SU$ with adj. tensors}\\
\downarrow \hspace{2.5cm} &\downarrow&\swarrow\\
SU(2) \hspace{2cm}&Spin(3)&\\
\end{eqnarray*}

Here one is able to actually do calculations by starting with the $Spin(10)$
model instead of trying to guess ones way up! We thus produced duals to
the $Spin$ groups listed above with a number of spinors corresponding to
1 spinor of $Spin(10)$. Since these theories are connected to
a variety of other interesting models, it would have been nice to generate
duals for those small groups with an arbitrary number of spinors, but
this proved to be very difficult, too. Our analysis nevertheless lead to
a better understanding of the problems of our generalization attempts,
thereby uncovering an interesting phenomenon.

{\bf In some duals models the full global symmetry is hidden on
the fundamental level on the magnetic side }

In addition, the self-consistency of the models obtained in this
way is further evidence for the correctness of Pouliot's duals.

First let's consider going from $Spin(10)$ to a smaller $Spin$
group. To achieve this, we give a vev to $k$ fundamental fields,
breaking the electric group to $Spin(10-k)$. Or in terms
of the gauge invariant polynomials, the meson $M^{i j} = Q^i Q^j$ gets a vev
of rank $k$.

The matter content of the electric theory is:
\begin{itemize}
\item $n-k$ vectors ($n$ was the number of $Spin(10)$ vectors)
\item a number of spinors and conjugate spinors corresponding to the
decomposition
of one $Spin(10)$ spinor
\item some singlets left over from the higgsing
($k(n-k)+\frac{k(k+1)}{2}$ in number).
\end{itemize}

The global rotation symmetry is $SU(n-k) \times SU(N_S) \times SU(N_C)$, where
$N_S$ and $N_C$ denote the number of spinors and conjugate spinors.

Next let's study the effect on the magnetic theory: the meson
is a magnetic gauge singlet, so its vev only enters through the superpotential:

\begin{eqnarray*}
W=\cdots +M qsq + \cdots & \leadsto & W=\cdots + <M> qsq +M qsq+ \cdots .\\
\end{eqnarray*}

Since $<M>$ only appears with this cubic combination, the effect is
simpler than in other examples, like SUSY QCD: whereas a
quadratic term would give mass to some fields, this new cubic term
leaves the theory almost unchanged. The only effect is that the
global $SU(n)$  symmetry is broken to a $SU(n-k) \times SO(k)$ subgroup.
The gauge group stays unchanged.
The fluctuations of the components of M that got a vev around
their expectation values correspond to the remaining singlets
on the electric side and one can get rid of them by writing down mass terms
on both sides.

At first sight the resulting dual pairs seem to be inconsistent, since the
global symmetries
are not the same:

{\bf Global Symmetries} (in addition to  $U(1)$ and $U(1)_R$)
\par\medskip
\begin{tabular}{|c|c||c|c|}
\hline
gauge gr.&$N_S+\tilde{N_S} \rule{0cm}{0.48cm}$& el. gl. symm.& mag. gl. symm.\\
\hline
\hline
$Spin(10)$&1&$SU(n)$&$SU(n)$\\
$Spin(9)$&1&$SU(n)$&$SU(n)$\\
$Spin(8)$&1 + 1&$SU(n)\times U(1)$&$SU(n) \times SO(2)$\\
$Spin(7)$&2&$SU(n) \times SU(2)$&$SU(n) \times SO(3)$\\
\hline
$Spin(6)$&2 + 2&$SU(n) \times SU(2) \times SU(2) \times U(1)$&$SU(n) \times
SO(4)$
\\
$Spin(5)$&4&$SU(n) \times SU(4)$&$SU(n) \times SO(5)$\\
$Spin(4)$&4 + 4&$SU(n) \times SU(4) \times SU(4) \times U(1)$&$SU(n) \times
SO(6)$
\\
$Spin(3)$&8&$SU(n) \times SU(8)$&$SU(n) \times SO(7)$\\
\hline
\end{tabular}
\par\medskip
(Here we used $n$ always as the number of electric fundamentals in the
theory under considerations, whereas above it was the number of
fundamentals in the $Spin(10)$ we started with).

For small $k$ some basic group isomorphisms seem to save the day, but starting
{}from $k=4$ ($Spin(6)$), where a $U(1)$ factor is missing on the magnetic
side, the magnetic global symmetry is always only a subgroup of the full
electric  symmetry.

Since these theories, nonetheless, \emph{should} be dual, we offer the
following explanation: on the magnetic side the full symmetry is not realized
linearly on the elementary fields; only part of it is. The missing generators
 act in a much more complicated way, exchanging fundamental fields and
composite fields. Only the effective theory in the far infrared has this
symmetry. Let us illustrate this in two examples chosen from the above table.
In each case, we will see
how the fundamental fields combine with composite gauge
invariant fields to form multiplets of the full global
symmetry. The electric gauge invariants are then identified with
those multiplets of invariants on the magnetic side.

\subsubsection{$Spin(5)$ with 4 Spinors}

Consider the following derived dual pair:

\begin{center}
\begin{tabular}{|c||c||c|c|c|c|}
\hline
&$Spin(5)$&$SU(n)$&$SU(4)$&$U(1)$&$U(1)_R$\\
\hline
\hline
$Q$&5&$n$&$1$&$-1$&$1-\frac{3}{n+2}$\\
\hline
$S$&4&1&$4$&$\frac{n}{2}$&$1-\frac{3}{n+2}$\\
\hline
\end{tabular}
\medskip\par
$\updownarrow$
\medskip\par
\begin{tabular}{|c||c||c|c|c|c|}
\hline
&$SU(n)$&$SU(n)$&$SO(5)=Sp(2)$&$U(1)$&$U(1)_R$\\
\hline
\hline
$q$&$\bar{n}$&$\bar{n}$&$1$&$1$&$\frac{3}{n+2}-\frac{1}{n}$\\
\hline
$q\prime$&$n$&1&$1$&$-n$&$-1+\frac{6}{n+2}+\frac{1}{n}$\\
\hline
$s$&$\Box \; \! \! \! \Box$ &$1$&$1$&0&$\frac{2}{n}$\\
\hline
$t$&$\bar{n}$&$1$&$5$&$0$&$1-\frac{1}{n}$\\
\hline
\hline
$Y$&$1$&$n$&$1$&$n-1$&$3-\frac{9}{n+2}$\\
\hline
$M$&$1$&$\Box \; \! \! \! \Box$ &$1$&$-2$&$2-\frac{6}{n+2}$\\
\hline
$N$&$1$&$1$&$5$&$n$&$2-\frac{6}{n+2}$\\
\hline
\end{tabular}
\medskip\par
with: $W= Mqsq+\det s+Yqq\prime+Ntq\prime +tst$.
\end{center}

We find the following mapping of gauge invariant operators:

Baryons:
\begin{eqnarray*}
 Q_a^{5-2i} W^i &\leftrightarrow& q^{n+2i-5} s^{n-3+i} W^{2-i} q\prime\\
 (SS)_s Q^3_a &\leftrightarrow& q^{n-3} t^3\\
(SS)_a  Q^4_a &\leftrightarrow& \left \{ q^{n-4} s^{n-2} W^2 + q^{n-4} t^4
\right \} \\
(SS)_a W^2 &\leftrightarrow& \left \{ {q^ns^n + W^2 N} \right \} \\
6&=& 1 +5\\
\end{eqnarray*}

Mesons:
\begin{eqnarray*}
 \begin{array}{ccc}
QQ &\leftrightarrow& M\\
{SS}_a &\leftrightarrow& \left \{ N +q^n \right \} \\
6&=& 1+5\\
 \end{array}
&\hspace{1cm}&
 \begin{array}{ccc}
(SS)_s Q^2_a &\leftrightarrow& q^{n-2} t^2 \\
(SS)_a Q &\leftrightarrow& \left \{ Y + t q^{n-1} \right \} \\
6&=&1+5\\
 \end{array}
\\
\end{eqnarray*}

The operators transforming under the asymmetric tensor representation (the 6)
of the global $SU(4)$ decompose as a $5+1$ under the $Sp(2)$
symmetry that is visible on the fundamental fields on the
magnetic side. Among the mesons, the $1$s are fundamental fields in the
lagrangian, whereas the $5$s are composite fields. Among the baryons, both
the $1$s and $5$s are composite, but they have a quite different structure.

Another interesting thing is to observe how the operators transforming
like a symmetric tensor (as a 10) under $SU(4)$ are realized on the magnetic
side (for example $(SS)_s Q^2_a$). As mentioned before, these were the most
problematic in trying
to generalize to more interesting dual pairs. Here the system avoids
the trouble in a rather tricky way: while on the electric side the 10 is a
symmetric combination of two 4s, on the magnetic side it is an antisymmetric
combination of two 5s, which can easily be  realized in terms of a magnetic
baryon.

\subsubsection{Spin(3) with 8 Spinors}

To see that the above construction is not special to the $Spin(5)$ case
we'll demonstrate that it also works in this derived duality. Again,
the full $SU(8)$ multiplet corresponding to a given electric invariant
is realized through combining elementary fields and composite invariants
transforming under an $SO(7)$ subgroup.

\begin{center}

\begin{tabular}{|c||c||c|c|c|c|}
\hline
&$Spin(3)$&$SU(n)$&$SU(8)$&$U(1)$&$U(1)_R$\\
\hline
\hline
$Q \rule{0cm}{.48cm}$&$3$&$n$&$1$&$-1$&$1-\frac{1}{n+2}$\\
\hline
$S \rule{0cm}{.48cm}$&$2$&1&$8$&$\frac{n}{2}$&$1-\frac{1}{n+2}$\\
\hline
\end{tabular}
\medskip\par
$\updownarrow$
\medskip\par
\begin{tabular}{|c||c||c|c|c|c|}
\hline
&$SU(n+2)$&$SU(n)$&$SO(7)$&$U(1)$&$U(1)_R$\\
\hline
\hline
$q \rule{0cm}{.48cm}$&$\overline{n+2}$&$\bar{n}$&$1$&$1$&$0$\\
\hline
$q\prime \rule{0cm}{.48cm}$&$n+2$&1&$1$&$-n$&$-1+\frac{3}{n+2}$\\
\hline
$s \rule{0cm}{.48cm}$&$\Box \; \! \! \! \Box$&$1$&$1$&0&$\frac{2}{n+2}$\\
\hline
$t \rule{0cm}{.48cm}$&$\overline{n+2}$&$1$&$7$&$0$&$1-\frac{1}{n+2}$\\
\hline
\hline
$Y \rule{0cm}{.48cm}$&$1$&$n$&$1$&$n-1$&$3-\frac{3}{n+2}$\\
\hline
$M \rule{0cm}{.48cm}$&$1$&$\Box \; \! \! \! \Box$ &$1$&$-2$&$2-\frac{2}{n+2}$\\
\hline
$N \rule{0cm}{.48cm}$&$1$&$1$&$7$&$n$&$2-\frac{2}{n+2}$\\
\hline
\end{tabular}
\medskip\par
with: $W= Mqsq+\det s+Yqq\prime+Ntq\prime +tst$.
\end{center}

Mesons:
\begin{eqnarray*}
 \begin{array}{ccc}
QQ &\leftrightarrow& M\\
{SS}_a &\leftrightarrow& \left \{ N +q^n t^2 \right \} \\
28&=&7 + 21\\
 \end{array}
&\hspace{1cm}&
 \begin{array}{ccc}
(SS)_s Q &\leftrightarrow&Y + q^{n-1} t^3 \\
36 &=& 1 + 35 \mbox{         }
 \end{array}
\\
\end{eqnarray*}

Again the symmetric combination, the 36, on the electric side is matched
with an antisymmetric combination of three 7s on the magnetic side.

In the same way, the other examples can be resolved. The obvious part of
the magnetic group for
$Spin(4)$ and $Spin(6)$ turns out to be the diagonal subgroup of the full
symmetry.

This type of construction will be encountered over and over again in what
follows. But already here we get a taste of what it means: at
no level of our analysis is it possible to generalize the obtained
theory to a higher number of spinors. Each of the subtle matchings depends
crucially on the number of flavors. The theories get more and
more complicated the further we move away from our starting point.
This behaviour is different from that encountered in SUSY QCD and the
other well known dualities, where perturbations drive us from one dual in a
series
of equally fundamental ones to another one from the same series.
The duality we
started with is a rather special case, not a part in a chain.
Going along flat directions produces new dual pairs because it has to, but
these still reflect the structure of the original dual, leading to
complicated
matter content and hidden global symmetries.
Duals for similar groups, like higher $Spin$ groups, $Spin$ with more
matter or exceptional groups have, in general, a similarly
complicated structure and it seems to be more a coincidence that in some
special cases the magnetic theory simplifies to the examples from above.
Still, they can be regarded as a hint of how complicated this
structure may look.

\subsection{Two Simple Duals}

In a similar way we can derive a dual for $SU(3)$ with $n$ flavors
in terms of an $SU$ gauge group with a symmetric tensor by
following the $G_2$ branch in the above diagram obtained by giving a vev to
the spinor of $Spin(7)$.
It is interesting since it is an example of a theory having two dual
descriptions in terms of a simple gauge group.
A number of theories with more than one dual are known \cite{pouasy,luty}, but
usually at least one of the possible dual descriptions involves a product
gauge group.

The electric theory is the well-known SUSY QCD with $N_C=3$, the first
dual is the original one constructed by Seiberg, the second dual is
the one derived from Pouliot's theory via $G_2$. This derivation involves
the same calculations as in the previous examples. Starting from the
dual of \cite{pou7} ($Spin(7)$ with $n$ spinors dual to $SU(n-4)$ with
a symmetric tensor and $n$ antifundamentals) one gives vev to a spinor of
$Spin(7)$ and then to a fundamental of the resulting $G_2$. The
corresponding magnetic fields are fundamental gauge singlets
multiplying cubic superpotential terms. Only the global symmetry
is broken, the gauge group and matter content of the magnetic theory stay the
same.

\begin{center}
\begin{tabular}{|c||c|c|c|c|}
\hline
$SU(3)$&$SU(n)_L$&$SU(n)_R$&$U(1)$&$U(1)_R$\\
\hline
\hline
$3 $&$n$&1&1&$1-\frac{3}{n}$\\
\hline
$\bar{3}$&1&$n$&$-1$&$1-\frac{3}{n}$\\
\hline
\end{tabular}
\medskip\par
$\updownarrow$
\begin{flushleft}
Seiberg:
\end{flushleft}
\begin{tabular}{|c||c||c|c|c|c|}
\hline
&$SU(n-3)$&$SU(n)_L$&$SU(n)_R$&$U(1)$&$U(1)_R$\\
\hline
\hline
$q \rule{0cm}{.48cm}$&$n-3$&$\bar{n}$&1&$-1+\frac{3}{n}$&$\frac{3}{n}$\\
\hline
$\bar{q}
\rule{0cm}{.48cm}$&$\overline{n-3}$&1&$\bar{n}$&$1-\frac{3}{n}$&$\frac{3}{n}$\\
\hline
$M \rule{0cm}{.48cm}$&$1$&$n$&$n$&0&$2-\frac{6}{n}$\\
\hline
\end{tabular}
\medskip\nopagebreak\par
with: $W= Mq\bar{q}$.\pagebreak[2]
\begin{flushleft}\
Derived from Pouliot:
\end{flushleft}\nopagebreak
\begin{tabular}{|c||c||c|c|c|}
\hline
&$SU(n-2)$&$SU(n)_D$&$U(1)$&$U(1)_R$\\
\hline
\hline
$q
\rule{0cm}{.48cm}$&$\overline{n-2}$&$\bar{n}$&$0$&$\frac{2}{n-2}-\frac{6}{n(n-2)
}$\\
\hline
$s \rule{0cm}{.48cm}$&$\Box \; \! \! \! \Box$&1&$0$&$\frac{2}{n-2}$\\
\hline
$t^{+} \rule{0cm}{.48cm}$&$\overline{n-2}$&$1$&3&$1-\frac{1}{n-2}$\\
\hline
$t^{-} \rule{0cm}{.48cm}$&$\overline{n-2}$&$1$&-3&$1-\frac{1}{n-2}$\\
\hline
\hline
$M \rule{0cm}{.48cm}$&$1$&$\Box \; \! \! \! \Box$ &$0$&$2-\frac{6}{n}$\\
\hline
\end{tabular}
\medskip\par
with: $W= Mqsq + \det s + q^{+} s q^{-}$.
\end{center}

Again, the second dual has only parts of the global symmetry visible on
the fundamental fields. As in the examples we considered before
different fundamental gauge singlets combine with composite fields
into multiplets of the full global symmetry. We get the following
mapping of gauge invariant operators in the two dual descriptions:

\begin{center}
\begin{tabular}{r|c|cc|cc}
&&Seiberg&$SU(n) \times SU(n)$&Pouliot&$SU(n)_{Diag.}$\\
\hline
Meson & $3 \bar{3}$ &$M$&$\tableau{1} \times \tableau{1}$&
$M+q^{n-2}$&$\tableau{2} + \tableau{1 1}$\\
Baryon & $3^3_a$ & $q^{n-3}$ &$\tableau{1 1 1} \times 1$& $q^{n-3}
t^{+}$&$\tableau{1 1 1}$\\
Antib. & $\bar{3}^3_a$ & $\bar{q}^{n-3}$ &$1 \times \tableau{1 1 1}$&
$q^{n-3} t^{-}$&$\tableau{1 1 1}$\\
\end{tabular}
\end{center}
Note that while in the original Seiberg dual all mesons become
fundamental fields which are singlets under the magnetic gauge group, in the
alternative dual description, the meson multiplet in magnetic theory
consists partly of fundamental and partly of composite fields.

\section{New Dualities}

Recently a dual description for $E_6$ with 6 fundamentals
has been found \cite{ramond}. The author shows that this theory is actually
self-dual.
Self-dual models deserve special attention since
it was argued in \cite{leigh} that they are closely related to
the existence of exactly marginal operators and hence to the existence of
fixed lines rather than fixed points.
One can get a condition for this to occur by studying
$\beta$ functions. The requirements that the $\beta$ functions
for all the couplings involved vanish, have to be linear dependent to
allow a line of solutions. With the analysis
of \cite{leigh} it is easy to see that this is associated with an exactly
marginal operator quadratic in the mesons appearing in the superpotential.
If we are
dealing with only one type of field and have only one
candidate operator to become a fundamental magnetic gauge singlet field,
this happens if the R-charge of the this meson becomes 1.
For $E_6$ this candidate operator is the 3-index symmetric composite
field and its R-charge becomes 1 for $N_F=6$.

 Similarly, we will argue that $E_7$ with 4 flavors is self-dual. By
following flat directions of those two theories and perturbing them with
mass terms we will construct some more new duals.
As evidence that we really constructed new dual theories, we take
that the 't Hooft anomaly matching conditions are satisfied
and
the fact that we can consistently flow from one to the other.
The fact that, perturbing a conjectured dual pair, we obtain a
self-consistent new dual pair is a highly non-trivial check of the
validity of the dual we started with.

It is much more difficult to check that the gauge invariant operators match
for the exceptional groups than for
the classical groups. Even though the basic invariant tensors are well known,
one can build higher invariant tensors out of the fundamental ones. There
exist several relations between contracted products of invariant tensors.
The real task is to find the independent ones. This problem is not yet
solved for the exceptional groups other than $G_2$.

Probably it is instructive to illustrate this problem with the solved $G_2$
example, following the lines of \cite{giddings}. $G_2$ has the symmetric
invariant $\delta^{\alpha \beta}$, the 7-index totally antisymmetric
$\epsilon$ tensor and and 3-index antisymmetric $f^{\alpha \beta \gamma}$.
Several relations between contracted products of those can be deduced from
Fierz identities of $Spin(7)$, since $G_2$ is a subgroup of $Spin(7)$, for
example
$$ f^{\alpha \beta \gamma} f^{\alpha \delta \epsilon}+f^{\alpha \delta
\gamma} f^{\alpha \beta \epsilon} = 2 \delta^{\beta \delta} \delta^{\gamma
\epsilon}-\delta^{\gamma \delta} \delta^{\beta \epsilon}- \delta^{\beta
\gamma} \delta^{ \delta \epsilon}$$
Only very few relations like this are known for the larger exceptional
groups. For $G_2$ the full set of those relations tells us that the only
independent
higher order invariant is $$\tilde{f}^{\alpha \beta \gamma \delta}=
\epsilon^{\alpha \beta \gamma \delta \epsilon \nu \rho} f_{\epsilon \nu
\rho}.$$
The full set of gauge invariant operators for $G_2$ is hence a 2-index
symmetric meson and 3-,4- and 7-index totally antisymmetric higher
composites.

\subsection{$E_6$ with 6 Flavors}

In \cite{ramond} Ramond found that $E_6$ with 6 flavors is self-dual.
The electric and magnetic fields transform under the global $SU(6) \times
U(1)_R
$ symmetry as follows:

\begin{center}
\begin{flushleft}
electric:
\end{flushleft}
\begin{tabular}{|c||c||c|c|}
\hline
&$E_6$&$SU(6)$&$U(1)_R$\\
\hline
\hline
$Q $&$27$&6&$\frac{1}{3}$\\
\hline
\end{tabular}
\begin{flushleft}
magnetic:
\end{flushleft}
\begin{tabular}{|c||c||c|c|}
\hline
&$E_6$&$SU(6)$&$U(1)_R$\\
\hline
\hline
$q $&$27$&$\bar{6}$&$\frac{1}{3}$\\
\hline
$Z$&$1$&$\tableau{3}$&$1
$\\
\hline
\end{tabular}
\par\medskip
with: $W= Z q^3$.
\end{center}

As evidence, he showed that the 't Hooft anomaly matching conditions are
satisfied and constructed a matching of some of the gauge invariant operators
(as mentioned above, finding all independent gauge invariant operators of
$E_6$ is still an unsolved problem). The three flavor symmetric invariant
becomes a fundamental field in the magnetic
theory. The $N_F=6$ case is exactly the case, where this field gets R-charge 1.
Since this is the only invariant appearing as a fundamental gauge singlet
in the magnetic theory, this indicates a possible self-duality,
as mentioned before.

Ramond found that there exist at least one independent higher invariant,
the sixth order composite invariant transforming like $$({\bf
27},\tableau{1})^6\sim({\bf 1},\tableau{2 2 2})\ $$
This is mapped to the corresponding magnetic invariant, the flavor indices
raised with 2 $\epsilon$ tensors. The latter construction is only possible with
6 flavors.

\subsection{$E_7$ with 4 Flavors}

As in the $E_6$ case from \cite{ramond}, the dual theory of
$E_7$ with 4 flavors is again $E_7$ with 4 flavors. The theory has a
global $SU(4) \times U(1)_R$ symmetry, the electric fields transforming
as

\begin{center}
\begin{tabular}{|c||c||c|c|}
\hline
&$E_7$&$SU(4)$&$U(1)_R$\\
\hline
\hline
$Q $&$56$&4&$\frac{1}{4}$\\
\hline
\end{tabular}

\end{center}

\noindent
and the magnetic fields as

\begin{center}

\begin{tabular}{|c||c||c|c|}
\hline
&$E_7$&$SU(4)$&$U(1)_R$\\
\hline
\hline
$q $&$56$&$\bar{4}$&$\frac{1}{4}$\\
\hline
$M$&$1$&$\tableau{4}$&$1$\\
\hline
\end{tabular}

\end{center}
\noindent
In addition the magnetic theory has a superpotential $W=M q^4$.

The global symmetries satisfy the 't Hooft anomaly matching conditions.
$E_7$ has as invariant tensors an 2-index antisymmetric tensor
$f^{\alpha \beta}$ and a 4-index symmetric invariant
$d^{\alpha \beta \gamma \delta}$ \cite{giddings}. While the
gauge invariant polynomial associated with $d^{\alpha \beta \gamma \delta}$
gets
 mapped to the gauge singlet M in the magnetic theory,
the antisymmetric invariant gets mapped to the corresponding
invariant of the magnetic theory:

\begin{eqnarray*}
Q^i_{\alpha} Q^j_{\beta} f^{\alpha \beta} &\leftrightarrow&
\epsilon^{ijkl} q_{k,\alpha} q_{l,\beta} f^{\alpha \beta} \\
\end{eqnarray*}

(with $i,j,k,l=1,\dots,4$ being flavor indices).
\noindent
M is again the only operator mapped to an elementary gauge singlet, and has
R-charge 1, which we already met as an indication of self-duality.

For some of the following duals to be consistent there must exist a 6th order
invariant transforming like:

$$({\bf 56},\tableau{1})^6\sim({\bf 1},\tableau{3 3})\ .$$

Under the duality, this would be mapped to the corresponding magnetic
invariant, the
flavor indices this time raised with 3 $\epsilon$ tensors.

This matching again is rather special to the 4 flavor
case. As with all of the exceptional group duals, one cannot dial
the number of colors in the magnetic theory to compensate
for changing the number of flavors in the electric theory.

\subsection{Derived Dualities}
\subsubsection{Going Along Flat Directions}

An interesting thing to do is to go along the flat directions of
this $E_7$ theory and the $E_6$ theory of \cite{ramond} and study the resulting
dual pairs.

As a first step we study the flat directions associated with those gauge
invariant
polynomials appearing as fundamental fields on the magnetic side.
(the 4-index symmetric tensor M in $E_7$ and the 3-index symmetric tensor Z
in $
E_6$)

The electric theory gets higgsed to:

$\begin{array}{ccccccccc}
E_7 & \rightarrow & E_6 &\rightarrow& F_4 & \rightarrow & Spin(8) &
\rightarrow & \ldots \\
4 \cdot 56&& 3\cdot 27 + 3 \cdot \bar{27} && 5 \cdot 26 && 4\cdot 8_V
+ 4 \cdot 8_S + 4 \cdot 8_C\\
\end{array}$

and:

$\begin{array}{ccccccc}
E_6 &\rightarrow& F_4 & \rightarrow & Spin(8) &
\rightarrow & \ldots \\
6\cdot 27  && 5 \cdot 26 && 4\cdot 8_V
+ 4 \cdot 8_S + 4 \cdot 8_C\\
\end{array}$

and some additional singlet fields.

On the magnetic side the field that gets a vev is a fundamental gauge singlet.
It only affects the dynamics through the superpotential. It acts in a very
simple
way: the gauge group stays unchanged, no fields become
massive, only the global symmetry gets broken to the subgroup leaving
$<M> q^4$ ($<Z> q^3$) in the $E_7$ ($E_6$) theory invariant.
This is exactly the same mechanism that was at work in the theories studied
in Section 3.

After giving mass to the singlet fields that remain after higgsing, one
gets the following dual pairs:

{\bf $F_4$ with 5 fund. and $E_6$:}

This dual is obtained by giving a vev to the 3-index invariant of $E_6$ (Z),
breaking the electric theory to $F_4$:

\begin{center}
\begin{tabular}{|c||c||c|c|}
\hline
&$F_4$&$SU(5)$&$U(1)_R$\\
\hline
\hline
$Q$&$26$&$5$&$\frac{2}{5}$\\
\hline
\end{tabular}
\par\medskip
$\updownarrow$
\par\medskip
\begin{tabular}{|c||c||c|c|}
\hline
&$E_6$&$SU(5)$&$U(1)_R$\\
\hline
\hline
$q$&$27$&$\bar{5}$&$\frac{4}{15}$\\
\hline
$t$&$27$&$1$&$\frac{2}{3}$\\
\hline
\hline
$Z$&$1$&$\tableau{3}$&$\frac{6}{5}$\\
\hline
$X$&$1$&$\tableau{2}$&$\frac{4}{5}$\\
\hline
\end{tabular}
\par\medskip
with: $W= t^3 + Xtq^2 +Z q^3$.
\end{center}

The magnetic gauge singlets Z and X correspond to the electric gauge singlets
built with the gauge invariant tensors $d^{\alpha \beta \gamma}$ and
$\delta^{\alpha \beta}$. In addition, there are 6th order magnetic composite
invariants $q^6$, $t q^5$, $t^2 q^4$ and the remaining 3rd order $t^2 q$.

After raising flavor indices, these should correspond to electric singlets
transforming like:

$$( \tableau{2 2} , \tableau{2 2 1} , \tableau{2 2 2} , \tableau{1 1 1 1}
)\ .$$

We are hence led to conjecture, that these correspond to higher order
invariants of $F_4$.

{\bf $Spin(8)$ with 4/4/4 and $E_6$:}

{}From the above dual we can go down another step to $Spin(8)$ by giving a vev
to a component of the 3-index invariant of $F_4$ (Z):

\begin{center}

\begin{tabular}{|c||c||c|c|c|c|}
\hline
&$Spin(8)$&$SU(4)_V$&$SU(4)_S$&$SU(4)_C$&$U(1)_R$\\
\hline
\hline
$Q$&$8_V$&$4$&$1$&$1$&$\frac{1}{2}$\\
\hline
$S$&$8_S$&$1$&$4$&$1$&$\frac{1}{2}$\\
\hline
$C$&$8_C$&$1$&$1$&$4$&$\frac{1}{2}$\\
\hline
\end{tabular}
\par\medskip
$\updownarrow$
\par\medskip
\begin{tabular}{|c||c||c|c|}
\hline
&$E_6$&$SU(4)_{Diag.}$&$U(1)_R$\\
\hline
\hline
$q$&$27$&$\bar{4}$&$\frac{1}{6}$\\
\hline
$t$&$27$&$1$&$\frac{2}{3}$\\
\hline
$v$&$27$&$1$&$\frac{2}{3}$\\
\hline
\hline
$Z$&$1$&$\tableau{3}$&$\frac{3}{2}$\\
\hline
$X$&$1$&$\tableau{2}$&$1$\\
\hline
$Y$&$1$&$\tableau{2}$&$1$\\
\hline
\end{tabular}
\par\medskip
with: $W= t^3 +v^3 +  Xtq^2 + Yvq^2 + Z q^3$.
\end{center}

Here only the diagonal subgroup of the global symmetry is visible on the
fundamental
fields. The gauge singlets combine with composite fields into multiplets of
the full symmetry.

\begin{center}
\begin{tabular}{c|c|c|c}
el. invariant& $SU(4)^3$&
mag. invariant& $SU(4)_{Diag.}$ \\
\hline
$QSC$&$\tableau{1} \times \tableau{1} \times \tableau{1}$&
$Z+tq^5+vq^5+tvq$& $ \tableau{3} +2 \cdot \tableau{2 1} + \tableau{1 1 1} $ \\
$QQ$&$\tableau{2} \times 1 \times 1 $&$X$&$\tableau{2}$\\
$SS$&$1 \times \tableau{2}  \times 1 $&$Y$&$\tableau{2}$\\
$CC$&$1 \times 1 \times \tableau{2} $&$q^6$&$\tableau{2}$\\
\end{tabular}
\end{center}

{\bf $E_6$ with 3/3 and $E_7$}

This dual is obtained by starting with the $E_7$ theory from above
and giving a vev to the 4-index invariant (M):

\begin{center}

\begin{tabular}{|c||c||c|c|c|}
\hline
&$E_6$&$SU(3)_L$&$SU(3)_R$&$U(1)_R$\\
\hline
\hline
$Q$&$27$&$3$&$1$&$\frac{1}{3}$\\
\hline
$\bar{Q}$&$\bar{27}$&$1$&$3$&$\frac{1}{3}$\\
\hline
\end{tabular}
\par\medskip
$\updownarrow$
\par\medskip
\begin{tabular}{|c||c||c|c|}
\hline
&$E_7$&$SU(3)_{Diag.}$&$U(1)_R$\\
\hline
\hline
$q$&$56$&$\bar{3}$&$\frac{1}{6}$\\
\hline
$t$&$56$&$1$&$\frac{1}{2}$\\
\hline
\hline
$M$&$1$&$\tableau{4}$&$\frac{4}{3}$\\
\hline
$Z$&$1$&$\tableau{3}$&$1$\\
\hline
$X$&$1$&$\tableau{2}$&$\frac{2}{3}$\\
\hline
$Y$&$1$&$\tableau{1 1}$&$\frac{5}{3}$\\
\hline
\end{tabular}
\par\nopagebreak\medskip
with: $W= Mq^4+Zq^3t+Xq^2t^2+Yq^2$.
\end{center}

Again only the diagonal subgroup of the flavor symmetry is manifest in the
Lagrangian.

\begin{center}
\begin{tabular}{c|c|c|c}
el. invariant& $SU(3) \times SU(3)$ & mag. invariant &
 $SU(3)_{Diag.}$\\
\hline
$Q \bar{Q}$&$\tableau{1} \times \tableau{1} $&
$X+tq$& $ \tableau{2}  + \tableau{1 1} $ \\
$Q^2  \bar{Q}^2$&$\tableau{2} \times \tableau{2} $&$M+ \cdots
$&$\tableau{4} +\cdots$\\
$Q^3$&$\tableau{3}  \times 1 $&$Z$&$\tableau{3}$\\
$\bar{Q}^3$&$1 \times \tableau{3}$&$q^3 t^3$&$\tableau{3}$\\
\end{tabular}
\end{center}

In addition the magnetic invariants $Y$ and $q t^3$ seem to correspond to
an 5th order composite $E_6$ invariant involving fundamentals and
antifundamentals, totally antisymmetric under both flavor symmetries.

In a similar fashion, one should get an $E_6$ dual of $Spin(10)$  and $E_7$
duals
of $F_4$ and $Spin(8)$.

\subsubsection{Mass Perturbations}
Even though
 the two duals we started with do not allow mass terms,
most of the new pairs we derived in the previous subsection do.
While this time the effect on the electric side is really simple
(the mass term just removes one field from the low energy
effective action), the magnetic case gets rather complicated.
The analysis for the $F_4$ case is sketched in the Appendix.

\subsection{Summary and Interpretation}

In this section we started with two self-dual models: $E_6$ with 6 flavors
found in \cite{ramond} and
$E_7$ with 4 flavors.
{}From the first one we got dual descriptions for $F_4$ with 5 flavors and
$Spin(8)$ with 4 vectors, 4 spinors and 4 conjugate spinors in terms of a
magnetic $E_6$ gauge group with 6 flavors via the Higgs mechanism. From the
second one we got a dual for $E_6$ with 3 fundamentals and 3
antifundamentals in terms of a $E_7$ gauge theory with 4 flavors. Finally
we studied a mass term in the $F_4$ theory with
5 flavors and obtained a dual for $F_4$ with 4 flavors with a magnetic
$Spin(9)$ group.

We consider the fact that the anomaly matching conditions are satisfied,
the gauge invariant operators match and that perturbations drive us to
new consistent theories as enough evidence, that these conjectured duals are
in fact correct. Nevertheless we encounter the same features as in our
discussion of the dual pairs derived from $Spin(10)$: the duals become more
and more complicated the further we go from our starting pair, all the
constructions depend crucially on the number of flavors, only a subgroup of
the global symmetry group is classically-realized in the magnetic theory. It is
far from obvious how to generalize to higher number of flavors.

\section{Conclusions}

We gave a detailed discussion about some new duals derived from known
dual pairs. We found that often only a subgroup of the global
symmetry is linearly-realized on  the fundamental fields
of the magnetic theory. The
fundamental fields combine with composite fields into
multiplets of the full symmetry.
We studied self-dual models and applied the method
of \cite{leigh} to find candidates for self-dual models.
With this we presented some duals for very special matter content
of $E_6$ and $E_7$ theories, some of which are new. In the same spirit
one would expect $E_8$ with 2 adjoints to be self-dual, but we haven't
checked this conjecture yet.

While we were finishing this work we received the very interesting
paper \cite{new} where the authors also found these
accidental symmetries. Our results overlap with their work.

\appendix
\section*{Appendix: Mass Terms for $F_4$}

We start from the $F_4$ theory with 5 fundamentals.
With the additional mass term
for a fundamental the magnetic superpotential becomes

\begin{eqnarray*}
W=t^3+X t q^2 + Zq^3 +mX .\\
\end{eqnarray*}
The equation of motion for X  reads:
\begin{eqnarray*}
<tq^2> + m &=& 0 .\\
\end{eqnarray*}
Hence the operator $tq^2$ gets a vev, breaking the magnetic $E_6$ to
$Spin(9)$. After integrating out the heavy fields we arrive at the following
dual pair:

\begin{center}
\begin{tabular}{|c||c||c|c|}
\hline
&$F_4$&$SU(4)$&$U(1)_R$\\
\hline
\hline
$Q$&$26$&$4$&$\frac{1}{4}$\\
\hline
\end{tabular}
\par\medskip
$\updownarrow$
\par\medskip
\begin{tabular}{|c||c||c|c|}
\hline
&$Spin(9)$&$SU(4)$&$U(1)_R$\\
\hline
\hline
$s$&$16$&$\bar{4}$&$\frac{1}{4}$\\
\hline
$q$&$9$&$\bar{4}$&$\frac{3}{4}$\\
\hline
\hline
$Z$&$1$&$\Box\! \Box\! \Box$&$\frac{3}{2}$\\
\hline
$X$&$1$&$\Box\! \Box$&$\frac{1}{2}$\\
\hline
$N$&$1$&$\Box\! \Box$&$\frac{3}{2}$\\
\hline
\end{tabular}
\par\medskip
with: $W= Zs^2 q + N s^2 + M q^2 + q^2 s^2.$
\end{center}

The magnetic gauge singlets X and Z are again easily identified as
the symmetric electric invariants built out of three and two fields.
The appearance of N is somewhat mysterious. It has the right
R-charge to correspond to an electric gauge invariant built
out of 6 fields, but its transformation under flavor
rotations seems to be rather strange. But since
the exact form of the higher invariants of $F_4$ is not known we can not rule
out this possibility.

Applying the same procedure once again one can arrive at a duality between
$F_4$ with 3 flavors and $G_2$ with 4 flavors. This can be easily retraced
to a duality between $E_6$ and $G_2$, both with 4 flavors. The latter is
known to be confining \cite{giddings}. As we are unable to recover the
exact magnetic superpotential,
we cannot extract any useful
information. But it is nice to see that again the 't Hooft matching conditions
are satisfied and the gauge invariant operators seem to work out, too.

Similarly one should be able to obtain a $Spin$ dual for $E_6$ with two
fundamentals and two antifundamentals by adding a mass term.

\end{document}